\documentstyle[11pt]{article}
\def\fnote#1#2{\begingroup\def\thefootnote{#1}\footnote{#2}\addtocounter
{footnote}{-1}\endgroup}
\def\BM#1{\mbox{\boldmath{$#1$}}}

\begin{document}

\hfill{UTTG-01-16 }

\vspace{36pt}

\begin{center}
{\large {\bf {What Happens in a Measurement?}}}

\vspace{36pt}
Steven Weinberg\fnote{*}{Electronic address:
weinberg@physics.utexas.edu}\\
{\em Theory Group, Department of Physics, University of
Texas\\
Austin, TX, 78712}

\vspace{30pt}

\noindent
{\bf Abstract}
\end{center}

\vspace{10pt}

\noindent
It is assumed that in a measurement the system  under study interacts with a macroscopic measuring apparatus, in such a way that the density matrix of the measured system  evolves according to the Lindblad equation.  Under an assumption of non-decreasing von Neumann entropy, conditions on the operators appearing in this equation are given that are necessary and sufficient for the late-time limit of the density matrix to take the form appropriate for a measurement. Where these conditions are satisfied, the Lindblad equation can be solved explicitly.  The probabilities appearing in the late-time limit of this general solution are found to agree with the Born rule, and are independent of the details of the operators in the Lindblad equation.

\vfill

\pagebreak

\begin{center}
I. INTRODUCTION
\end{center}		

According to the Copenhagen interpretation of quantum mechanics, during a complete measurement the initial density matrix $\rho_{\rm initial}$ undergoes a collapse 
\begin{equation}
\rho_{\rm initial}\mapsto \rho_{\rm final}=\sum_\alpha p_\alpha\Lambda_\alpha
\end{equation}
where $\Lambda_\alpha=|\alpha\rangle\langle\alpha|$ are projection operators onto the complete set of orthonormal eigenvectors $|\alpha\rangle$  of whatever is being measured, satisfying the usual conditions
\begin{equation}
\Lambda_\alpha\Lambda_\beta=\delta_{\alpha\beta}\Lambda_\alpha~~~~~~~\sum_\alpha\Lambda_\alpha=\BM{1}~~~~~~~{\rm Tr}\Lambda_\alpha=1~~~~~\Lambda_\alpha^\dagger=\Lambda_\alpha\;,
\end{equation}
and $p_\alpha$ are probabilities, given by the Born rule 
\begin{equation}
p_\alpha=\langle\alpha|\rho_{\rm initial}|\alpha\rangle={\rm Tr}\Big(\Lambda_\alpha \rho_{\rm initial}\Big)\;.
\end{equation}
In a closed system in ordinary quantum mechanics the state vector evolves unitarily and deterministically, so as well known the collapse (1) cannot occur if the initial density matrix $\rho_{\rm initial}$ describes a pure state (or an ensemble of fewer pure states than the number of terms in $\rho_{\rm final}$).  In the original formulation[1] of the Copenhagen interpretation it was simply accepted that the change in a system during measurement in principle  departs from quantum mechanics.  We will instead adopt the popular modern  view that  the Copenhagen interpretation refers to open systems in which the transition (1) is driven by the interaction of the microscopic system under study with a suitable environment, a macroscopic external measuring apparatus (which may include an observer) chosen to bring this transition about.

Of course, this view of the Copenhagen interpretation just pushes  the hard problems of interpreting quantum mechanics to a  larger scale.  We make no attempt to address these problems in the present paper, beyond noting the conjecture[2] that  the unitary evolution of microscopic systems is merely a very good approximation, while the density matrix of combined systems with macroscopic parts in general evolves rapidly  and non-unitarily, and in particular undergoes the collapse (1) during a measurement.   Although in this paper we are focusing on 
ordinary quantum mechanics in an open system,  most of our analysis applies equally to closed systems in modified versions of quantum mechanics.

Whether in  open systems in ordinary quantum mechanics or in closed systems in some modified version of quantum mechanics,  in order to avoid instantaneous communication at a distance in entangled states, it is important to require that the density matrix at one time depends  on the density matrix at any earlier time, but not otherwise on  the state vector at the earlier time.[3].  This evolution can be linear but non-unitary in ordinary quantum mechanics if the system under study interacts with an environment that fluctuates randomly more rapidly than the rate at which the density matrix evolves (set by the interaction strength) if we average over these fluctuations.   We do not need to go into details regarding this interaction with the environment, because it is known that in the most general linear evolution that preserves the unit trace and Hermiticity of the density matrix and satisfies the condition of complete positivity[4], the density matrix satisfies the Lindblad equation[5]:
\begin{equation}
\frac{d\rho(t)}{dt}=-i\Big[{\cal H},\rho(t)\Big]+\sum_n \left(L_n\rho(t)L_n^\dagger-\frac{1}{2}L_n^\dagger L_n \rho(t)
-\frac{1}{2}\rho(t) L_n^\dagger L_n\right)\;,
\end{equation}
with constant matrices\fnote{**}{We limit the considerations of this paper to a Hilbert space of a finite dimensionality $d$.  Presumably they can be extended to infinite dimensional spaces, on which ${\cal H}$ and the $L_n$ act as suitably defined operators.} $L_n$ and ${\cal H}$.  
We then face three questions:
\begin{enumerate}
  \item What are the necessary conditions on the operators  $L_n$  and ${\cal H}$ for  the density matrix to approach a time-independent linear combination such as (1) of specific projection operators $\Lambda_\alpha$ at late times?
  \item  Are these conditions sufficient?  
  \item For such $L_n$  and ${\cal H}$, are the coefficients $p_\alpha$ of the $\Lambda_\alpha$ in this linear combination given by the Born rule (3)? 
 \end{enumerate}  
The answer to the first question is given in Section II, under the assumption that the $L_n$ satisfy the necessary and sufficient condition[6] that the von Neumann entropy $-{\rm Tr}(\rho\ln\rho)$ should never decrease:
\begin{equation}
\sum_n L_n^\dagger L_n=\sum_n L_n L^\dagger_n\;.
\end{equation}
The second and third questions are answered in Section III, where we give a general solution of the Lindblad equation, under the conditions found in Section II.

\begin{center}
II. NECESSARY CONDITIONS FOR A MEASUREMENT
\end{center}

First, let us consider some general aspects of the late-time behavior of the solutions of the Lindblad equation (4), without yet specializing to $L_n$ satisfying Eq.~(5).  Because Eq.~(4) is linear with time-independent coefficients, it has solutions that are generically of the form
\begin{equation}
\rho(t)=\sum_k v_k \exp\Big(\lambda_k t\Big)\;,
\end{equation}
where $v_k$ and $\lambda_k$ are the eigenmatrices and eigenvalues of the operator ${\cal L}$ in Eq.~(4):
\begin{equation}
{\cal L}v_k=\lambda_k v_k\;,
\end{equation}
\begin{equation}
{\cal L}v\equiv -i\Big[{\cal H},v\Big]+\sum_n \left(L_n\,v\,L_n^\dagger-\frac{1}{2}L_n^\dagger L_n\, v
-\frac{1}{2}v\, L_n^\dagger L_n\right)\;,
\end{equation}
with the normalization of each $v_k$ in Eq.~(6) of course depending on initial conditions.
(It is only  for the non-degenerate case that the solution of Eq.~(4) necessarily takes the form (6); if an eigenvalue $\lambda_k$ has an $N$-fold degeneracy, then $\exp(\lambda_kt)$ may be accompanied with a polynomial in $t$ of order up to $N-1$.)  Because ${\cal L}$ is in general not Hermitian the eigenvalues may be complex, and  the individual $v_k$ need not be Hermitian or positive, though the sum (6) must be both Hermitian and positive, 

Even if there are  eigenvalues $\lambda_k$ with positive-definite real parts, such terms cannot contribute to the sum (6).  If they did contribute then the sum of such terms would dominate $\rho(t)$  at late times. But ${\rm Tr}\rho(t)$ must remain constant, so the sum of terms with ${\rm Re}\lambda_k>0$ would have to be traceless.  Also, $\rho(t)$ must remain Hermitian and positive, so the sum of terms with ${\rm Re}\lambda_k>0$ would have to be Hermitian and positive. But then the eigenvalues of this sum would have to be real and positive and add up to zero, which is impossible unless all the eigenvalues vanish, in which case the sum vanishes.  The  same argument rules out any contribution of powers of time for any eigenvalues with ${\rm Re}\lambda_k=0$.  So we conclude that the asymptotic behavior of $\rho(t)$ is dominated by the sum of $v_k\exp(\lambda_k t)$ over all eigenmatrices with ${\rm Re}\lambda_k=0$, if there are any.  

In fact, as required by the constancy of the trace, there always is at least one eigenmatrix with $\lambda_k=0$.  We can think of ${\cal L}$ as a $d^2\times d^2$ matrix, acting on the space of $d\times d$ matrices.  Because Eq.~(4) preserves the trace of $\rho$, the unit $d\times d$ matrix ${\BM 1}$ is a left eigenvector of ${\cal L}$ with eigenvalue zero, so ${\rm Det}{\cal L}=0$, and therefore ${\cal L}$ also has a right eigenvector (not necessarily the unit matrix) with eigenvalue zero.  But in general there may be several  $v_k$ with $\lambda_k=0$.

In order to separate the real and imaginary parts of general eigenvalues, let us consider the quantity
\begin{equation}
{\rm Tr}\Big(v_k^\dagger v_k\Big)\,\lambda_k= {\rm Tr}\Big(v_k^\dagger {\cal L}v_k\Big)\;.
\end{equation}
A straightforward calculation gives
\begin{eqnarray}
&&{\rm Tr}\Big(v_k^\dagger v_k\Big){\rm Re}\lambda_k=-\frac{1}{2}{\rm Tr}\left(\sum_n [v_k\,,\,L_n^\dagger]^\dagger[v_k\,,\,L_n^\dagger]\right)\nonumber\\&&
~~~~~-\frac{1}{2}{\rm Tr}\left(v_k v_k^\dagger \sum_n\Big(L_n^\dagger L_n-L_n L_n^\dagger\Big)\right)\\
&&{\rm Tr}\Big(v_k^\dagger v_k\Big){\rm Im}\lambda_k=-{\rm Tr}\Big(v_k^\dagger [{\cal H},v_k]\Big)
+{\rm Im}{\rm Tr}\sum_n L_nv_k^\dagger[v_k,L_n^\dagger ]\end{eqnarray}
(See Appendix A.)  
It is difficult to make further progress without invoking some assumption that limits the nature of the $L_n$.  As mentioned in Sec. I, we shall assume that the $L_n$ satisfy the necessary and sufficient condition (5) for non-decreasing entropy.  In this case, Eq.~(10) simplifies to 
\begin{equation}
{\rm Tr}\Big(v_k^\dagger v_k\Big){\rm Re}\lambda_k=-\frac{1}{2}{\rm Tr}\left(\sum_n [v_k\,,\,L_n^\dagger]^\dagger[v_k\,,\,L_n^\dagger]\right)\;.
\end{equation}
We see immediately that  the real parts of all $\lambda_k$ are negative or zero.  The behavior of $\rho(t)$ for $t\rightarrow\infty$ is then dominated by the modes $v_k$ for which ${\rm Re}\lambda_k=0$, for which according to Eq. (12) $v_k$ must commute with all $L^\dagger_n$, and hence with all $L_n$.  (By taking the adjoint of Eq.~(7) we see that if $v_k$ is an eigenmatrix of ${\cal L}$ then so is $v_k^\dagger$, which must appear in (6) along with $v_k$ to keep $\rho$ Hermitian.  The adjoint of the condition that $v_k^\dagger$ commutes with $L_n^\dagger$ tells us that $v_k$ must commute with $L_n$.)  

Also, for such modes Eq.~(11) gives
\begin{equation}{\rm Tr}\Big(v_k^\dagger v_k\Big){\rm Im}\lambda_k=-{\rm Tr}\Big(v_k^\dagger [{\cal H},v_k]\Big)\;.
\end{equation}
But it must not be thought that a $v_k$ that commutes with all $L_n$ is necessarily an eigenmatrix of ${\cal L}$ with the real part of the eigenvalue zero and its imaginary part of first order in ${\cal H}$.    With $L_n$ subject to Eq.~(5) and $v_k$ commuting with all $L_n$, the eigenvalue equation (7) becomes
$$\lambda_k v_k=-i[{\cal H},v_k]\;,$$
but this is impossible if the space of matrices that commute with all $L_n$ is not invariant under commutation with ${\cal H}$.  Otherwise the commutator with ${\cal H}$ in Eq.~(7) will mix the eigenmatrices $v_k$ that commute with all $L_n$ with other matrices that do not commute with some ${L_n}$, giving an eigenvalue with negative-definite real part, whose contribution vanishes for $t\rightarrow\infty$.  

Here is an example. Take $d=2$, with a single $L_n$ given by $L=\ell\sigma_3$ (which trivially satisfies Eq.~(5)), and ${\cal H}=h\sigma_1$, with $h$ real.  This $L$ commutes with the projection operators $(1\pm \sigma_3)/2$, as required in a measurement of $\sigma_3$, but for $h\neq 0$ the commutator of ${\cal H}$ with these projection operators does not commute with them, so the measurment doesn't work.  We can see this in the late-time behavior of the solutions of the Lindblad equation.   In general the eigenmatrices  of ${\cal L}$ are $v_0\propto\BM{1}$, with eigenvalue zero; $v_{1}\propto\sigma_1$, with eigenvalue $-2|\ell|^2<0$; and two mixtures of $\sigma_2$ and $\sigma_3$, with eigenvalues $-|\ell|^2\pm(|\ell|^4-4h^2)^{1/2}$.  For $h=0$ there are two eigenmatrices with eigenvalue zero, which can be taken as $\BM{1}$ and $\sigma_3$, or equivalently as the projection matrices $(\BM{1}+\sigma_3)/2$ and       $(\BM{1}-\sigma_3)/2$ as needed in a measurement of $\sigma_3$; the other eigenmatrices both have eigenvalues $-2|\ell|^2$, corresponding to modes that disappear for $t\rightarrow \infty$.  On the other hand the late-time behavior of the density matrix is entirely different if $h$ is non-zero, though arbitrarily small.  In this case all eigenvalues have negative-definite real part, except the eigenvalue $\lambda_0=0$ associated with $v_0\propto\BM{1}$, and  for $t\rightarrow \infty$ the density matrix approaches the maximum entropy matrix $\BM{1}/2$, for which all probabilities are the same.

With this background, let us now consider what happens in a measurement.  We suppose that the microscopic system under study interacts with a macroscopic measuring apparatus, in such a way that the density matrix of the microscopic system evolves according to the Lindblad equation (4), with the measuring apparatus chosen so that the matrices $L_n$ and ${\cal H}$ have whatever properties are needed so that $\rho(t)$ at late times approaches a linear combination of projection operators $\Lambda_\alpha$ on the eigenstates $|\alpha\rangle$ of whatever is being measured.  As we have seen in Eq.~(12), in order for this to be the case without putting any constraints on the initial conditions that determine the coefficients in this linear combination, it is necessary that the matrices $L_n$ should commute with any linear combination of the $\Lambda_\alpha$, and hence with each 
$\Lambda_\alpha$:
\begin{equation}[L_n,\Lambda_\alpha]=0\;,          \end{equation}
from which it follows immediately that each $L_n$ must itself be a linear combination of the $\Lambda_\alpha$:
\begin{equation}
 L_n=\sum_\alpha \ell_{n\alpha}\Lambda_\alpha\;,
\end{equation}
with coefficients $\ell_{n\alpha}$ that are in general complex numbers.
(From Eq.~(14) it follows that  the eigenstates $|\alpha\rangle$  satisfying $\Lambda_\beta|\alpha\rangle=\delta_{\alpha\beta}|\alpha\rangle$ must be eigenstates of the $L_n$:
$$ L_n|\alpha\rangle=L_n\Lambda_\alpha|\alpha\rangle=\Lambda_\alpha L_n|\alpha\rangle=|\alpha\rangle \langle \alpha|L_n|\alpha\rangle$$
Then $L_n$ has the same action on any $|\alpha\rangle$ as does the sum (15) with $\ell_{n\alpha}= \langle \alpha|L_n|\alpha\rangle$, and since the $|\alpha\rangle$ form a complete set, $L_n$ must equal the sum (15).)  From Eq.~(15) the condition (5) for non-decreasing entropy follows trivially.

This leaves us with the matrix ${\cal H}$.  As remarked earlier, in order that the limiting behavior of $\rho(t)$ for general initial conditions should be a linear combination of the $\Lambda_\alpha$, it is necessary that the space of such linear combinations should be invariant under commutation with ${\cal H}$:
$$
[{\cal H},\Lambda_\alpha]=\sum_\beta h_{\alpha\beta}\Lambda_\beta\;.
$$
By multiplying this commutator on both the left and right with any $\Lambda_\beta$, we see that $0=h_{\alpha\beta}$, and therefore ${\cal H}$ must commute with all $\Lambda_\alpha$.  By the same argument used above for the $L_n$, we see then that ${\cal H}$ must be a linear combination of the $\Lambda_\alpha$:
\begin{equation}
{\cal H}=\sum_\alpha h_\alpha\Lambda_\alpha\;,
\end{equation}
with real coefficients $h_\alpha$.

This is a good place to bring up a complication.  The late-time behavior (1) is expected only for a {\em complete} measurement.  It is more common for measurements to be incomplete, in the sense that they do not lead to definite states $|\alpha\rangle$ with definite probabilities, but to equivalence classes of  states that are not distinguished by the measurement.  For instance, in a system consisting of two spins $1/2$, we might measure only the first spin, leaving the other undisturbed.  The states then fall into two classes, labeled by the $z$-components of the two spins: one class consists of $|1/2,1/2\rangle$ and  $|1/2,-1/2\rangle$, and the other consists of $|-1/2,1/2\rangle$ and  $|-1/2,-1/2\rangle$.  In incomplete measurements, instead of (1), the expected late-time limit of the density matrix is
\begin{equation}
\rho_{\rm initial}\mapsto\rho_{\rm final}=\sum_C \Lambda_C\rho_{\rm initial}\Lambda_C \;,
\end{equation}
where 
\begin{equation}
\Lambda_C=\sum_{\alpha\in C}\Lambda_\alpha\;,
\end{equation}
As far as the states within a single class are concerned, $\Lambda_C$ acts just like a unit matrix, so Eq,~(17) says that the measurement does nothing to what is not being measured.   For a complete measurement, where each  state belongs to a different class, Eq.~(17) reduces to Eqs.~(1) and (3).

 Eq.~(12) shows that in order for $\rho(t)$ to have some given asymptotic limit $\rho_{\rm final}$, it is necessary  for all $L_n$ to commute with 
this limit, and since this must be true for all $\rho_{\rm initial}$, the $L_n$ here must in particular commute with $\sum_C\Lambda_C\Lambda_\alpha\Lambda_C=\Lambda_\alpha$.  The same argument as given above for complete measurements then shows that, here too, eash $L_n$ must be a linear combination (15) of the $\Lambda_\alpha$.  Only now there is a constraint on the coefficients.  The commutator of the sum (15) with the limit (17) is 
$$ \Big[\sum_\alpha \ell_{n\alpha} \Lambda_\alpha\,,\,\sum_C \Lambda_C\rho_{\rm initial}\Lambda_C\Big]
=\sum_C\sum_{\beta,\gamma\in C}[\ell_{n\beta}-\ell_{n\gamma}]\Lambda_\beta\rho_{\rm initial}\Lambda_\gamma$$
which vanishes for all initial density matrices if $\ell_{n\beta}=\ell_{n\gamma}$ for all $\beta$ and $\gamma$ in the same class.  The same argument shows that $h_\beta=h_\gamma$ if $\beta$ and $\gamma$ are in the same class.
This is reasonable, because for an incomplete measurement the Lindblad equation must not distinguish between different states in the same class,  We will see in the next section that in this case 
the late-time limit of the density matrix does have the form (17).

\begin{center}
III. COLLAPSE OF THE DENSITY MATRIX
\end{center}

First let us give the solution of the Lindblad equation under the condition that the matrices $L_n$ and ${\cal H}$ in this equation are linear combinations (15), (16) of projection operators $\Lambda_\alpha$ satisfying Eq.~(2):
$$ 
L_n=\sum_{\alpha}\ell_{n\alpha}\Lambda_\alpha\;,~~~~~~~
{\cal H}=\sum_{\alpha}h_{\alpha}\Lambda_\alpha~~\;.$$
It is straightforward to check that Eq.~(4) is then satisfied by
\begin{equation}
\rho(t)=\sum_{\alpha\beta}\Lambda_\alpha M \Lambda_\beta\;\exp(\lambda_{\alpha\beta}t)\;,
\end{equation}
where
\begin{equation}
\lambda_{\alpha\beta}=-\frac{1}{2}\sum_n\Big|\ell_{n\alpha}-\ell_{n\beta}\Big|^2+i\;{\rm Im}\sum_n \ell_{n\alpha}\ell^*_{n\beta}
-i\Big(h_{\alpha}-h_{\beta}\Big)\;.
\end{equation}
and $M$ is an arbitrary  matrix, independent of $\alpha$, $\beta$, and time. [See Appendix B.]  To relate $M$ to the initial value of $\rho(t)$ at $t=0$, set $t=0$ in Eq.~(19) and use the completeness condition $\sum_\alpha\Lambda_\alpha=\BM{1}$.  We see that $\rho(0)=M$, and so
\begin{equation}
\rho(t)=\sum_{\alpha\beta}\Lambda_\alpha \rho(0) \Lambda_\beta\;\exp(\lambda_{\alpha\beta}t)\;,
\end{equation}
This is our general solution.[7]

Now consider the behavior of this solution at late times.  The only terms in the sum (21) that do not decay exponentially are those with $\ell_{n\alpha}=\ell_{n\beta}$ for all $n$.  
If for the moment we rule out degeneracy,  so that $\ell_{n\alpha}$ can equal $\ell_{n\beta}$ for all $n$ only for $\alpha=\beta$, then all $\lambda_{\alpha\beta}$ have negative-definite real part except those with $\alpha=\beta$, for which the imaginary as well as the real parts of $\lambda_{\alpha\alpha}$ vanish.  These terms then dominate the asymptotic behavior[8]  of the density matrix for $t\rightarrow \infty$:
\begin{equation}
\rho(t)\rightarrow \sum_\alpha \Lambda_\alpha\rho(0)\Lambda_\alpha=\sum_\alpha \Lambda_\alpha \langle \alpha|\rho(0)|\alpha\rangle
\end{equation}
This is just the behavior (1) called for by the Copenhagen interpretation, with probabilities $p_\alpha$ given by the Born rule (3).  

The case of degeneracy arises in an incomplete measurement, in which  we only measure whether the system is in some state or other in a class  of states that are not distinguished by the measurement.  As indicated at the end of the previous section, in this case we expect $\ell_{n\alpha}$ to equal $\ell_{n\beta}$ for all $n$ and $h_\beta=h_\gamma$ if (and only if) $|\alpha\rangle$ and $|\beta\rangle$ are in the same class.  Then Eq.~(21) has the expected late-time behavior (17).

It is striking that although the detailed time-dependence of the density matrix depends on the coefficients $\ell_{n\alpha}$ and $h_\alpha$ appearing in the matrices in the Lindblad equation, the asymptotic limit for $t\rightarrow\infty$ for both complete and incomplete measurements does not depend on these details, depending only on the initial condition $\rho(0)$ and on what it is that  is being measured.  
This, of course, is just what we require of a measurement.

\vspace{20pt}

\begin{center}
Acknowledgments
\end{center}

I am grateful for correspondence with P. Pearle, and regarding the condition for non-decreasing entropy, with H. Narnhofer,  D. Reeb, and R. Werner.  This material is based upon work supported by the National Science Foundation under Grant Number PHY-1316033 and with support from The Robert A. Welch Foundation, Grant No. F-0014.

\vspace{10pt}

\begin{center}
{APPENDIX A: Derivation of Eqs.~(10) and (11) }
\end{center}
\renewcommand{\theequation}{A.\arabic{equation}}
\setcounter{equation}{0}

We start with the desired result, and work back to the problem it solves.  For a general matrix $v$, consider the quantity 
\begin{eqnarray}
&&R\equiv-\frac{1}{2}{\rm Tr}\left(\sum_n [v\,,\,L_n^\dagger]^\dagger[v\,,\,L_n^\dagger]\right)-\frac{1}{2}{\rm Tr}\left(v v^\dagger \sum_n\Big(L_n^\dagger L_n-L_n L_n^\dagger\Big)\right)\nonumber\\
&&+i{\rm Im}{\rm Tr}\sum_n L_nv^\dagger[v,L_n^\dagger ]-i{\rm Tr}\Big(v^\dagger [{\cal H},v]\Big)
\end{eqnarray}
Expanding each term, this is
\begin{eqnarray}
&&R=-\frac{1}{2}{\rm Tr}\sum_n L_n v^\dagger v\,L_n^\dagger
+\frac{1}{2}{\rm Tr}\sum_n  v^\dagger L_n v\,L_n^\dagger
+\frac{1}{2}{\rm Tr}\sum_n L_n v^\dagger L_n^\dagger\,v
-\frac{1}{2}{\rm Tr}\sum_n v^\dagger L_n  \,L_n^\dagger\,v\nonumber\\&&
-\frac{1}{2}{\rm Tr}\sum_n v v^\dagger  L_n^\dagger L_n
+\frac{1}{2}{\rm Tr} \sum_n v v^\dagger  L_n L^\dagger_n\nonumber\\
&&+\frac{1}{2}{\rm Tr}\sum_n v^\dagger L_n v  L_n^\dagger-\frac{1}{2}{\rm Tr}\sum_n L_n v^\dagger L_n^\dagger v\nonumber\\&&
-i{\rm Tr}\Big(v^\dagger [{\cal H},v]\Big)\;.
\end{eqnarray}
The third and eighth terms cancel; the fourth and sixth terms cancel; the second and seventh terms add to give the term
${\rm Tr} v^\dagger\sum_n L_n v L_n^\dagger$ in ${\rm Tr}v^\dagger{\cal L}v$;  the first and fifth terms give the terms
$-{\rm Tr} v^\dagger v \sum_n  L_n^\dagger L_n /2$ and $-{\rm Tr} v^\dagger\sum_n L^\dagger_n  L_n v/2$ in ${\rm Tr}v^\dagger{\cal L}v$; and the last term gives the Hamiltonian term in ${\rm Tr}v^\dagger{\cal L}v$.  We conclude that 
\begin{equation}
R={\rm Tr}\Big(v^\dagger{\cal L}v\Big)\;.
\end{equation}
The first two terms in (A.1) are real, while the last two are imaginary, so
\begin{equation}
{\rm Re}{\rm Tr}\,v^\dagger{\cal L}v=-\frac{1}{2}{\rm Tr}\left(\sum_n [v\,,\,L_n^\dagger]^\dagger[v\,,\,L_n^\dagger]\right)-\frac{1}{2}{\rm Tr}\left(v v^\dagger \sum_n\Big(L_n^\dagger L_n-L_n L_n^\dagger\Big)\right)
\end{equation}
\begin{equation}
{\rm Im}\,{\rm Tr}\,v^\dagger{\cal L}v={\rm Im}{\rm Tr}\sum_n L_nv^\dagger[v,L_n^\dagger ]\,-{\rm Tr}\Big(v^\dagger [{\cal H},v]\Big)
\end{equation}
Taking $v$ to be one of the eigenmatrices $v_k$ of ${\cal L}$, with ${\cal L}v_k=\lambda_kv_k$ then gives Eqs.~(10) and (11).

\vspace{10pt}

\begin{center}
{APPENDIX B: Derivation of Eqs. (19) and (20) }
\end{center}
\renewcommand{\theequation}{B.\arabic{equation}}
\setcounter{equation}{0}

We try a solution of the Lindblad equation
\begin{equation}
\rho(t)=\sum_{\alpha\beta}\Lambda_\alpha M \Lambda_\beta f_{\alpha\beta}(t)\;.
\end{equation}
With $\Lambda_\alpha$ and ${\cal H}$ given by Eqs.~(15) and (16), the Lindblad equation (4) becomes
\begin{equation}`\sum_{\alpha\beta}\Lambda_\alpha M \Lambda_\beta\frac{d}{dt} f_{\alpha\beta}(t)=\sum_{\alpha\beta}\lambda_{\alpha\beta}\Lambda_\alpha M \Lambda_\beta f_{\alpha\beta}(t)\;,
\end{equation}`
where 
\begin{equation}
\lambda_{\alpha\beta}=C_{\alpha\beta}-\frac{1}{2}C_{\alpha\alpha}-\frac{1}{2}C_{\beta\beta}-i(h_\alpha-h_\beta)
\end{equation}
and
\begin{equation}
C_{\alpha\beta}=\sum_n \ell_{n\alpha}\ell_{n\beta}^*\;.
\end{equation}
This has an obvious solution of the same form as (19):
\begin{equation}
f_{\alpha\beta}(t)=\exp\Big(\lambda_{\alpha\beta}t\Big)f_{\alpha\beta}(0)\;.
\end{equation}
To get a more useful expression for $\lambda_{\alpha\beta}$, we note that 
$$-\frac{1}{2}\left|\ell_{\alpha}-\ell_{\beta}\right|^2=-\frac{1}{2}C_{\alpha\alpha}-\frac{1}{2}C_{\beta\beta}+{\rm Re}\sum_n \ell_{n\alpha}\ell_{n\beta}^*=C_{\alpha\beta}-\frac{1}{2}C_{\alpha\alpha}-\frac{1}{2}C_{\beta\beta}-i{\rm Im}\sum_n \ell_{n\alpha}\ell_{n\beta}^* $$
so Eq.~(B.3) is the same as Eq.~(20).

\vspace{10pt}

\begin{center}
{\bf ---------}
\end{center}

\vspace{10pt}

\begin{enumerate}

\item N. Bohr, Nature {\bf 121}, 580 (1928).
\item G. C. Ghirardi, A. Rimini,
and T. Weber, Phys. Rev. D {\bf 34}, 470 (1986);  P. Pearle, Phys. Rev. A {\bf 39}, 2277 (1989), and in {\em Quantum Theory: A Two-Time Success Story} (Yakir Aharonov Festschrift), eds. D. C. Struppa \& J. M. Tollakson (Springer, 2013), Chapter 9. [arXiv:1209.5082]
\item N. Gisin, Helv. Phys. Acta {\bf 62}, 363 (1989); Phys. Lett. A {\bf 143}, 1 (1990).  This is discussed in a wider context by J. Polchinski, Phys. Rev. Lett. {\bf 66}, 397 (1991).
\item  If any entangled density matrix for a compound system ${\cal S}\otimes {\cal S}$ consisting of two isolated copies of a system ${\cal S}$ remains positive for a  range of future times if it is positive at an initial time, then the linear mapping $\rho(t)\rightarrow\rho(t')$ of the density matrix of ${\cal S}$  for $t'>t$ in this range is completely positive,  as shown by F. Benatti, R. Floreanini, and R. Romano,  J. Phys. A Math. Gen.  {\bf 35}, L551 (2002).  For complete positivity see W. F. Stinnespring, Proc. Am. Math. Soc. {\bf 6}, 211 (1955); M. D. Choi, J. Canada Math. {\bf 24}, 520 (1972).  For its  implications, see  M. D. Choi, {\it Linear Algebra and its Applications} {\bf 10}, 285 (1975).  
\item G. Lindblad, Commun. Math. Phys. {\bf 48}, 119 (1976); V. Gorini, A. Kossakowski and E. C. G. Sudarshan, J. Math. Phys. {\bf 17}, 821 (1976).  For a straightforward derivation, see P. Pearle, Eur. J. Phys. {\bf 33}, 805 (2012).
\item F. Benatti and R. Narnhofer, Lett. Math. Phys. {\bf 15}, 325 (1988).  (Their result, which applies for infinite as well as finite Hilbert spaces, takes the form of an inequality.  When limited to finite Hilbert spaces, it iw equivalent to  the equality (5).)  It was earlier shown by T. Banks, M. Peskin, and L. Susskind, Nuclear Phys. B {\bf 244}, 125 (1984), that a sufficient (though not necessary) condition for non-decreasing entropy is that the $L_n$ are Hermitian.  Of course, if the $L_n$ are Hermitian then Eq.~(5) is automatically satisfied.
\item This solution was  given in the second edition of S. Weinberg, {\em Lectures on Quantum Mechanics} (Cambridge University Press, Cambridge, UK, 2015), Section 6.9, for the special case where all $L_n$ are Hermitian.
\item This behavior is seen in several of the examples presented by Pearle in ref. [5].
\end{enumerate}

  \end{document}